\documentclass[prb,aps]{revtex4}

\usepackage{epsfig}
\usepackage{verbatim}
\usepackage{graphicx}
\usepackage{caption}
\usepackage{amsmath}
\usepackage{epstopdf}
\usepackage{float}

\usepackage{url}

\usepackage{braket}

\usepackage{hyperref}

\usepackage{multirow}

\graphicspath{./}

\def\zero{S$_{\bf 0}$}
\def\five{S$_{\bf 5}$}
\def\sixty{S$_{\bf 60}$}
\def\seventy{S$_{\bf 70}$}
\def\deg{^{\circ}}

\begin{document}

\title{Band Splitting in Bilayer Stanene Electronic Structure scrutinized via first principle DFT Calculations}
\author{Mohammed Ghadiyali}
\email{ghadiyali.mohd@physics.mu.ac.in}
\author{Sajeev Chacko}
\email{sajeev.chacko@physics.mu.ac.in; sajeev.chacko@gmail.com}
\affiliation{Department of Physics, University of Mumbai, Kalina Campus, Santacruz (E), Mumbai - 400 098, India.}
\

\begin{abstract}

The recent work on stanene as quantum spin Hall insulators made us investigate bilayer stanene using first principle
calculations.
With an aim of improving and developing new properties, via modulating the stacking order (and angle) of the bilayers.
This stacking of layers has been proven technique for modulating the properties of monolayer materials.
Here we design multiple bilayer systems, with different stacking angles and AA and AB configurations.
Rather observing an improvement in bandgap due to spin-orbit coupling (SOC), we witness a splitting of the band due to
SOC, a characteristic behavior of stacked MoS$_2$ sheets.
This splitting of the bands gives rise to different, independent and distinct spin-up and spin-down channels, manifesting a
valley dependent spin polarization.
Also, as a contrast to stacked MoS$_2$ system we notice in our system the stacking angle and order, does effect
electronic states.

\end{abstract}

\maketitle

%...............................................................
%...............................................................

\section{Introduction}
The exfoliation of graphene~\cite{Graphene} has generated a huge interest in the field of condensed matter physics.
The honeycomb lattice structure of graphene exhibits necessary features required by the Haldane model~\cite{Haldane_Model}.
Materials represented by Haldane model generally host quantum Hall effect.
The spin-orbital coupling~(SOC) in such materials could lead to quantum spin Hall effect
(QSH)~\cite{Topological_Insulator_Hassan}.
The first experimental observation of QSH in HgTe quantum well~\cite{HgTe_Quantum_Well} led to the development of
topological insulators~(TI)~\cite{Topological_Insulator_Hassan}.
TIs are the class of material which could assist to bring the dream of creating dissipationless conduction channel for
electrons a reality.
Two-dimensional versions of these TIs are called as quantum spin Hall insulators (QSHI)~\cite{100_Conducting}.
However, some of the 2D QSHI have a weak SOC, a very small bandgap ($\sim 10^{-3}$ meV), making it unsuitable for
most of the applications~\cite{Graphene_Spin_Hall}.
Much of the work in the last decade has been dedicated to identification and synthesis of these materials.

One way around is to investigate other honeycomb lattices such as those formed out of group IVA elements right below
Carbon in the periodic table.
This has led to buckled honeycomb lattice of Silicon and Germanium, which have been computationally
verified~\cite{2D_Si_Ge_Computational,2D_Si_Ge_Computational_2} and experimentally
synthesized~\cite{2D_Si_Exprimental,2D_Ge_Exprimental}.
Unfortunately, these also have a very small bandgap\cite{Si_Ge_band_gap,2D_Si_Ge_Computational}.
A 2D buckled honeycomb lattice of Tin (stanene), which is computationally predicted~\cite{Tin_Spin_Hall, Tin_Spin_Hall_2}
as a QSHI, is supposed to have a sufficient  bandgap required for practical use as QSHI.
It was fabricated on Bi$_2$Te$_3$ $(111)$ substrate via molecular beam epitaxy~
\cite{Stanene_Synthesis_China,Stanene_Synthesis_India}.
The issue with this is that although it has been synthesized, its electronic states are pinned on the substrates $i.e.$
stanene's valance bands with conduction bands of the substrate, giving stanene a metallic interface and losing its QSHI
nature.
This unwanted pinning of conducting states have been recently overcome by synthesizing stanene on PbTe substrate,
allowing the stanene to retain the bandgap\cite{Stanene_Synthesis_China_2}.
Apart form this recently synthesis of planar stanene has been reported on Ag$_2$Sn
substrate~\cite{Stanene_Synthesis_non_buckled}.
The bandgap of free-standing stanene is deemed to be sufficient but still small.
To improve the bandgap, functionalization as well as passivation with different groups like -H
(hydrogenation)~\cite{Stanene_H}, -F (halogen)~\cite{Tin_Spin_Hall, Tin_Spin_Hall_2}, -CN (Cyanide)~\cite{Stanene_CN}
etc. have been proposed.
Studies of van der Waal type structures like stanene on graphene~\cite{Stanene_on_graphene}, stanene between two Ba
layer~\cite{Stanene_between_Ba},  have been reported improve the bandgap.
Apart from its use as QSHI, stanene also has been predicted to be a very good material to adsorb CO, NO$_X$, SO$_2$,
etc. gases~\cite{Stanene_As_Gas_Absorber}.
Also, recently stanene's mechanical property~\cite{Stanene_mechanical} and it applicability in devices~\cite{Stanene_device} have been studied, implying its importance.
It may be noted that most of the above-mentioned work is \textit{in silico}.

In this paper, we perform the first-principle calculations of the bilayer of 2D tin (stanene), stacked similarly to that of the
bilayer graphene.
A similar approach had been used by Kan \textit{et al.}, for performing density functional theory~(DFT) studies on the two
lattice of stanene, which are stacked upon each other like AB or Bernal-stacked form.
In their study, the bilayer structure had lost its buckling.
However, it has been reported that this structure has a lower energy than free-standing stanene making it more stable and
having a higher bandgap~\cite{Bilayer_Stanene}.
We use a similar approach to construct the stanene bilayer keeping the structure and mechanical exfoliation of bilayer
graphene as our template.
In addition to varying interlayer spacing, we also varied the stacking angles as discussed in section~\ref{sec:method}.
Interestingly, instead of exhibiting a QSHI nature, the proposed systems displayed distinctive attributes of band structure
due to band splitting are observed.

%...............................................................

\section{Method}
\label{sec:method}

First principle plane waves pseudopotential calculations were performed using the DFT method as implemented in
Quantum ESPRESSO (QE) package~\cite{QE}.
We have used Perdew-Burke-Ernzerhof (PBE)~\cite{PBE} exchange-correlation functional and included the DFT-D2 van der
Waals corrections~\cite{DFT-D_1, DFT-D_2} for representing the electron dispersion effect due to the bilayer.
The kinetic energy cutoff was converged at 50 Ry, with a k-point mesh (Monkhorst-Pack) of $18\times 18\times 1$ .
To minimize the interaction between neighboring layers, a vacuum $\approx 20$\AA~was set above and below the bilayer.
To include the effect of SOC we used fully relativistic pseudopotentials obtained from QE pseudopotential repository
website~\cite{QE_Psudo}.
The calculation of the band structure and the partial charge densities were performed both with and without SOC~\cite{Hybrid_no_use}.
The k-point path for band structure calculations was set according to high symmetry path in Brillouin zone for hexagonal
lattices, as $M\rightarrow \Gamma\rightarrow K\rightarrow M$.
The required k-points for this path were generated using the XCrysDens~\cite{Xcrystal} package.

A bilayer structure can be constructed by three types of rotations of one
of its layer with respect to the second one: (1)~relative rotation between
two layers within a single unit cell, (2)~rotation one of the entire layer with
respect to a reference axis and (3)~combination of these two cases. 
The 2nd and 3rd methods of rotation have been widely used for modification of
electronic states in bilayer graphene, boron-nitrate-graphene heterostructures
etc. Here a sheet of single layer composed of large number of unit cells is
rotated with respect to the second layer.
This generates a large number of atoms and electrons in the system.
Such a system is extremely difficult to study using the DFT method used in the
present work.
Generally, tight binding method is preferred for these systems.
Further, one of the reasons for the modification of electronic states by
interlayer rotation is to break the PT-symmetry, which we could observe
with the 1st type rotation for the systems under consideration.
Hence, we have carried out the construction of the bilayer stanene using only
with the first rotation method.

\begin{figure}[H]
	\centering
	\includegraphics[scale=0.5]{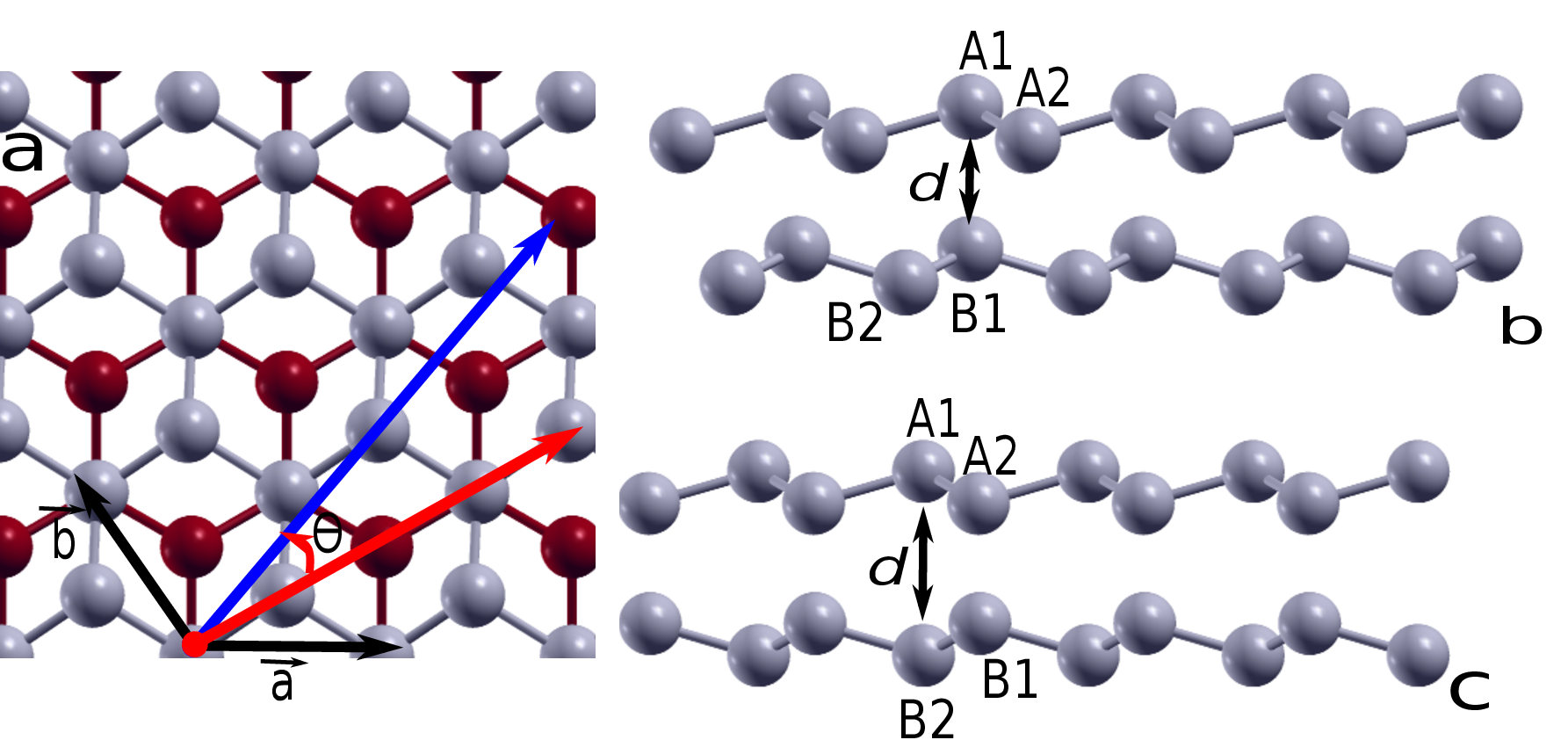}
	\caption{(color online)
		    (a)~Structure of bilayer stanene. The gray spheres represent atoms in the top layer and the red spheres represent atoms the bottom layer.
		    The angle, $\theta$, between layers is defined as shown in the image.
		    % The value of $\theta$ is changed as $0^{\circ}$, $5^{\circ}$, $60^{\circ}$ and $70^{\circ}$.
			The top and bottom atoms of the top layer are labeled as A1 and A2, respectively, while those in the lower plane are labeled as B1 and B2, respectively.
			(b)~The AA configuration where the A1 atoms are on top of the B1 atoms.
			(c)~The AB configuration where the A1 atoms are on top of the B2 atoms.
			These labels of the configurations (AA and AB) are strictly in the context of this paper.}
	\label{F1}
\end{figure}

The construction of the bilayer structures were performed by stacking the two
layers of stanene on each other with an interlayer angle $\theta$ as illustrated
in figure~\ref{F1}(a).
We then changed the angle of top layer with respect to the bottom layer from
$0^{\circ}$ to $90^{\circ}$ in steps of 5$^{\circ}$.
However, since these transformations were performed on single unit cell, only
0$^{\circ}$, 5$^{\circ}$, 60$^{\circ}$ and 70$^{\circ}$ angles bilayers could be
constructed.
For other stacking angles, the resultant structures were found to be
discontinuous and/or significantly distorted (see figure SI:1).
Therefore, only the un-distorted structures were selected for further analyzes.
We label these structures as: \zero, \five, \sixty~and \seventy, respectively.
This process was carried out both configurations of the tin bilayer as described in section~\ref{sec:results}.
Next, we varied the interlayer spacing~($d$) and computed the total energy as a function of $d$.
The structure with lowest energy was selected for the further calculations.
Choice of this approach over the conventional geometry optimization method was preferred due to the reason that a bilayer
or more layered structure can be mechanically exfoliated from their free-standing monolayer counterpart.
These mechanically exfoliated bilayers can be redesigned by simply rotating the layers with respect to each other.
This process gives an extra degree of freedom for development of new materials with unique properties.
The structures for present work were created using the VNL-Builder (version 2016.3)~\cite{VNL}.

%...............................................................

%...............................................................

\section{Result and Discussion}
\label{sec:results}

A free-standing stanene monolayer has a buckled honeycomb lattice with two atoms per primitive unit cell. Each of the tin
atoms is sp$^3$ hybridized to all its neighboring atoms.
It is also known to exhibit two stable configurations based on the extent of buckling.~\cite{Rivero_PRB_2014,
Broek_2Dmat_2014, Matusalem_PRB_2014}.
The optimized structure has a lattice constant of 4.67~\AA~with a Sn-Sn bondlength of 2.83~\AA~and a buckling height of
0.86~\AA.
The energetically less favored low-buckled structure has been investigated more extensively due to its exotic electronic
properties.

For the bilayer, a unit cell of a low-buckled structure having the following parameters was used: $a = b =
4.7$~\AA~\cite{Bilayer_Stanene} and $c~=~27$~\AA.
The intralayer Sn-Sn bondlength was same as that of the monolayer.
An initial separation between the layers was taken to be 2.4~\AA.
A interlayer spacing of 2.9~\AA~is reported for their geometry optimized planner bilayer stanene configuration by  Kan
\textit{et al.}
However, to obtain various configurations and especially the lowest energy structures, we start with a lower interlayer
spacing of 2.4~\AA.
Bilayer stanene inherently has two different geometry due to buckling.
To understand the structures, we have labeled the atoms in the top sheet as A1 and A2, representing the top and bottom
atom and, B1 and B2 the atoms in the bottom layer, as shown in figure~\ref{F1}(b).
If the atom A1 is on top of atom B1 than the system is labeled as AA configuration and if atom A1 is on top of atom B2
than the system is labeled as AB configuration.
The nomenclature of these structures is strictly in the context of this paper.

\begin{figure}[H]
	\centering
	\includegraphics[scale=0.5]{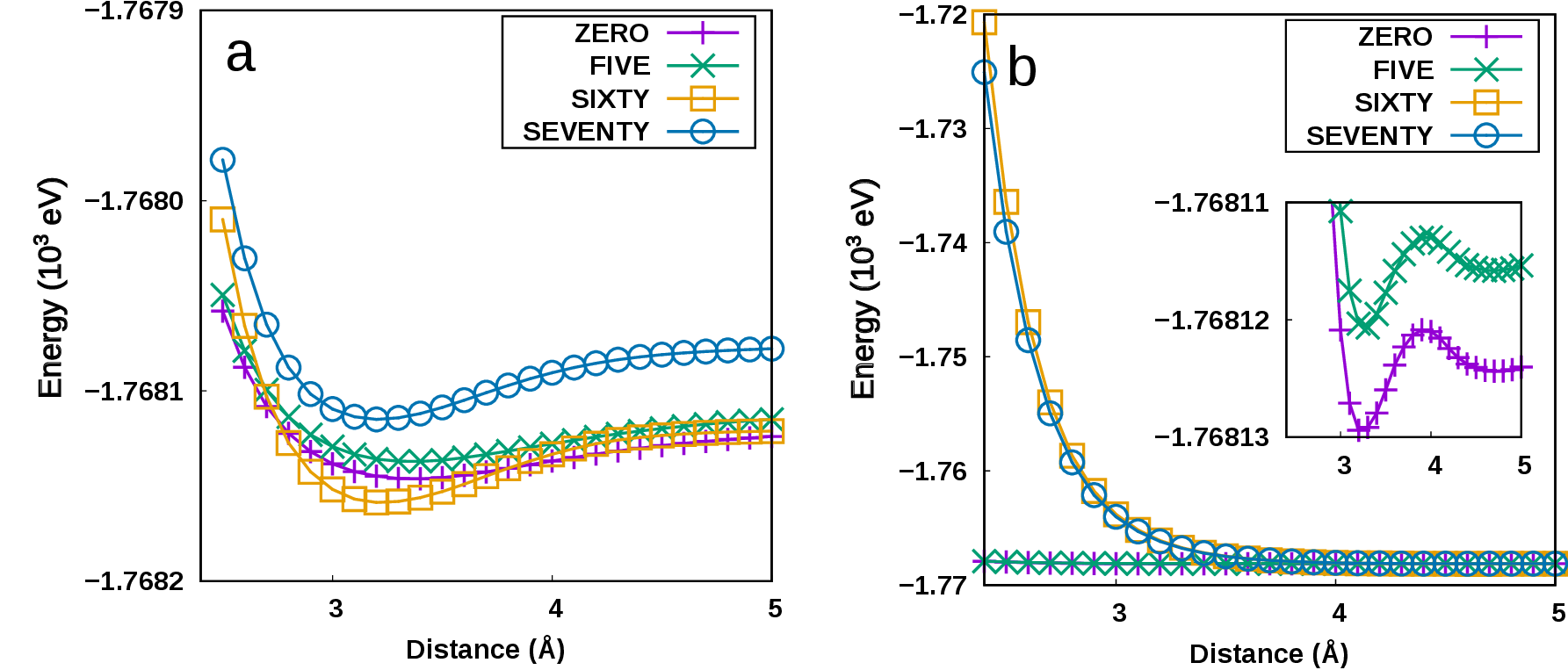}
	\caption{(color online) The total energy $v/s$ the interlayer spacing for the configurations (a) AA and (b) AB
         for interlayer angles of $0^{\circ}$, $5^{\circ}$, $60^{\circ}$ and $70^{\circ}$.
	 	The inset (b) is an enlarged  graph of plot for angles $0\deg$  and $5^{\circ}$ showing the second minima at
	 	$d=4.7$~\AA.}
	\label{fig:energy_distance_combine}
\end{figure}

For obtaining the minimum energy structure the interlayer spacing, $d$, was varied in the steps of 0.1~\AA, and the total
energy was computed for each structure.
The plot of energy versus interlayer spacing is given in figure~\ref{fig:energy_distance_combine}.
The spacing~$d$ corresponding to the lowest energy is given in table~\ref{T1}.
% These minimum energy structures were studied for further analyzes.
From table~\ref{T1} and the figure~\ref{fig:energy_distance_combine} only one minimum is observed for the
AA~configuration for all angles of rotation of the top layer with respect to the bottom layer.
On the other hand, for the AB configuration, two minima are observed for \zero~and \five~structures.
For reference check figure~\ref{fig:energy_distance_combine}.
Further, to determine if the proposed systems were energetically favorably, their cohesive energies per atom were calculated (see the supplementary information TABLE-I).
The cohesive energies of all the bilayer stanene structures studied are of the same order as that of the mono-layer stanene indicating that the proposed systems are energetically stable.
We then selected the structures with lowest energy for each of the four stacking angles for further analyzes.

%...............................................................

\begin{table}[H]
	\centering
	\begin{tabular}{|c|c|c|c|c|}
		\hline
		\multirow{2}{*}{Configuration} & \multicolumn{4}{c|}{d (\AA)}            \\ \cline{2-5}
		                                  & $S_{0}$  & $S_{5}$ & $S_{60}$   & $S_{70}$ \\ \hline
		AA                             & 3.4           & 3.4          & 3.2       														   & 3.2            \\ \hline
		AB                             & 3.2 (4.7) & 3.3 (4.7) &$\hspace{0.4cm}$4.6$\hspace{0.4cm}$& $\hspace{0.4cm}$4.6$\hspace{0.4cm}$ \\ \hline
	\end{tabular}
	\caption{Interlayer spacing~$d$ for different angles corresponding to the minimum
		energy for different configurations of the stanene bilayer. The values of $d$ for the second minimum for \zero and \five are given in the bracket.}
	\label{T1}
\end{table}

%...............................................................

%...............................................................

\begin{figure}[htbp]
	\centering
	\includegraphics[scale=0.6]{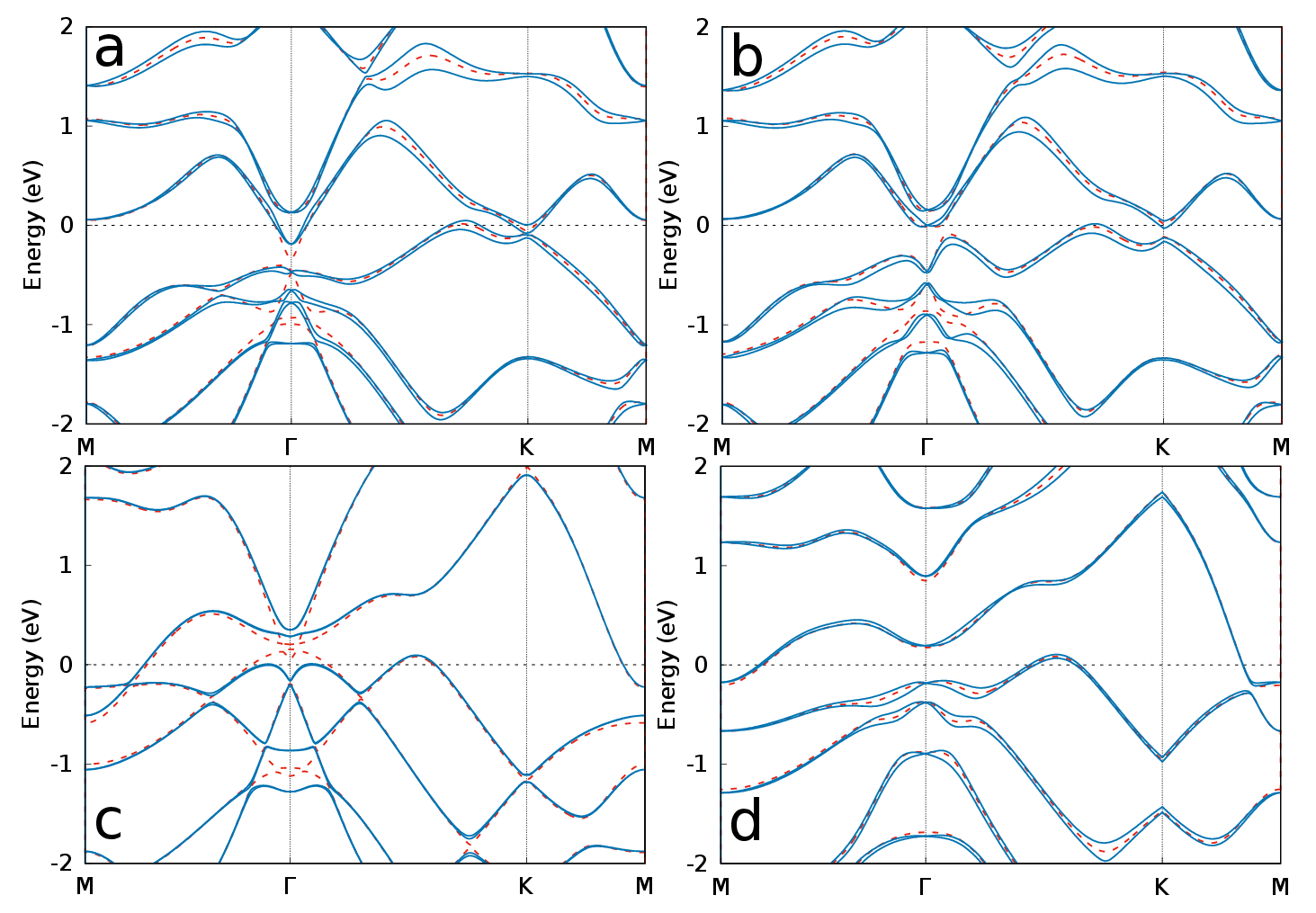}
	\caption{(color online) The band structures of the AA system, for angles  (a)~$0^{\circ}$, (b)~$5^{\circ}$,
			(c)~$60^{\circ}$ and (d)~$70^{\circ}$.
			The dotted red lines represent the bands without SOC and the blue lines represent the bands with SOC. The yellow translucent circles in (b) marks the position where due to band splitting the structure converts to metallic as described in the text. It is to be noted that Fermi Energy is set to be at zero on y-axis.}
	\label{all_aa_bands}
\end{figure}

\begin{figure}[H]
	\centering
	\includegraphics[scale=0.6]{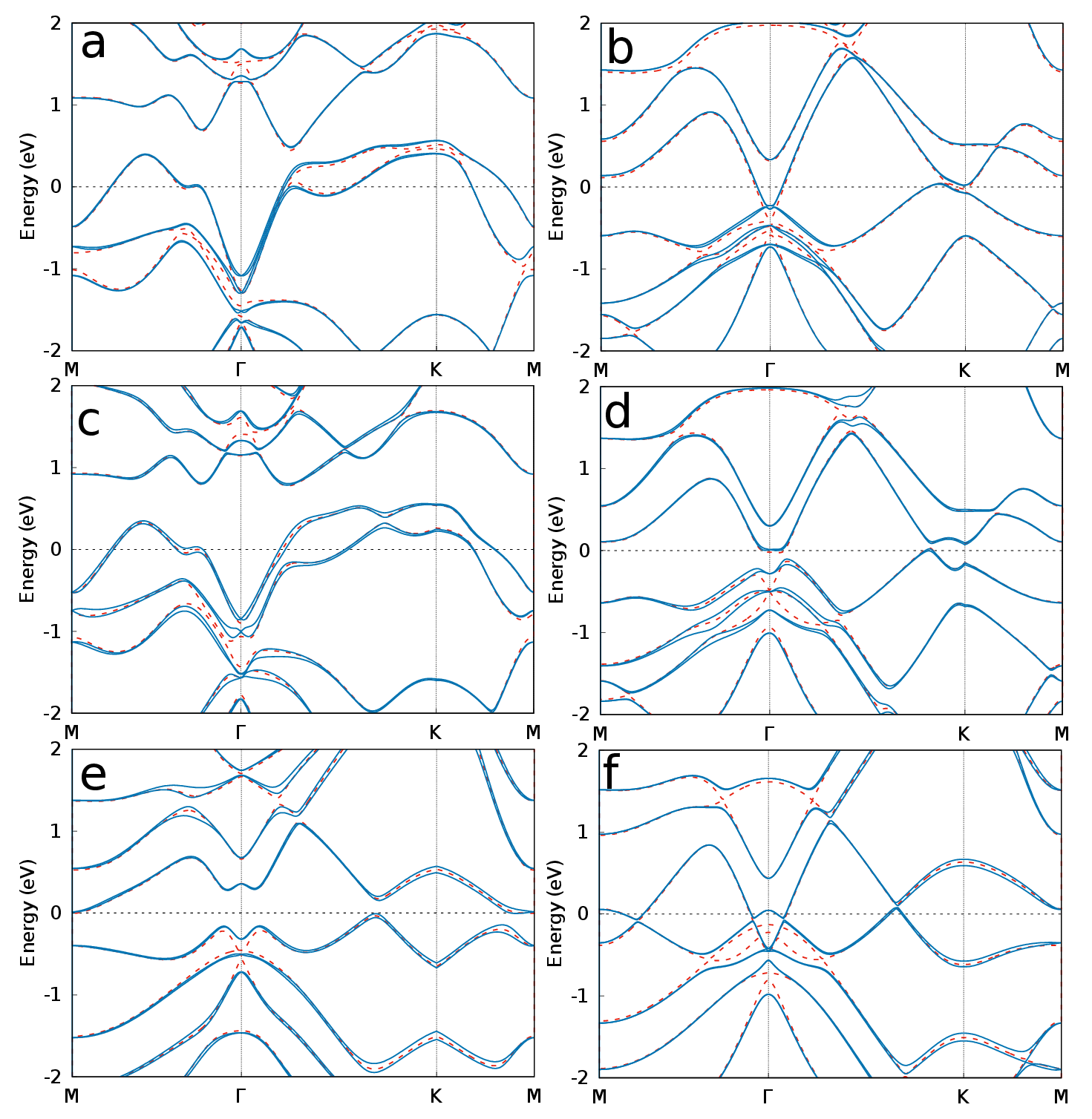}
	\caption{(color online) The band structures of the AB system for angles (a)~$0^{\circ}$ ($d=$3.2~\AA),
			(b)~$0^{\circ}$ ($d=$4.7~\AA), (c)~$5^{\circ}$ ($d=$3.3~\AA), (d)~$5^{\circ}$ ($d=$4.7~\AA), (f)~$60^{\circ}$
			and (e)~$70^{\circ}$. The dotted red lines represent the bands without SOC and the blue lines  represent the bands
			with SOC. The yellow translucent circles in (e) marks the position where due to band splitting the structure converts to metallic as described in the text. It is to be noted that Fermi Energy is set to be at zero on y-axis.} 
	\label{all_ab_bands}
\end{figure}

The band structures along the high symmetry points ($M\rightarrow\Gamma\rightarrow K\rightarrow M$) with and without
SOC are shown in figure~\ref{all_aa_bands} and~\ref{all_ab_bands} for configurations AA and AB, respectively.
It is clear from the band structures that SOC does not open up a gap (as expected in QSHI) but induces a band splitting.
Interestingly, in one of the case, this band splitting decisively converts the semimetallic nature of the system to a metallic
one while in other the semiconducting nature is converted into metallic.
Such a behavior for materials having Haldane model has not been reported.
As the conversation is due to band splitting via SOC, the split in bands could have different up and down states which can be
exploited for applications in valleytronics~\cite{Valleytronics}.
Hence, only the systems expressing this transition were further investigated, $i.e.$ the system with AA~configuration
having a stacking angle of $5^{\circ}$ (AA-\five) and AB~configuration having a stacking angle of $70^{\circ}$
(AB-\seventy).
In order to understand the nature of band splitting, a cue from the work of Zheng~\textit{et
al.}~\cite{MOS_SOC_SPLITING_1} and Zhu \textit{et al.}~\cite{MOS_SOC_SPLITING_2} on bilayer MoS$_2$ is taken as
the template.
The calculated results along with band decomposed charge density at specific k-points were investigated by the same
method employed by Zheng~\textit{et al.}~\cite{MOS_SOC_SPLITING_1} and Zhu \textit{et
al.}~\cite{MOS_SOC_SPLITING_2} due to the similarity of their and our systems in terms of geometric and electronic
structure.

%...............................................................

%...............................................................

%...............................................................

\begin{figure}[H]
	\centering
	\includegraphics[scale=1.0]{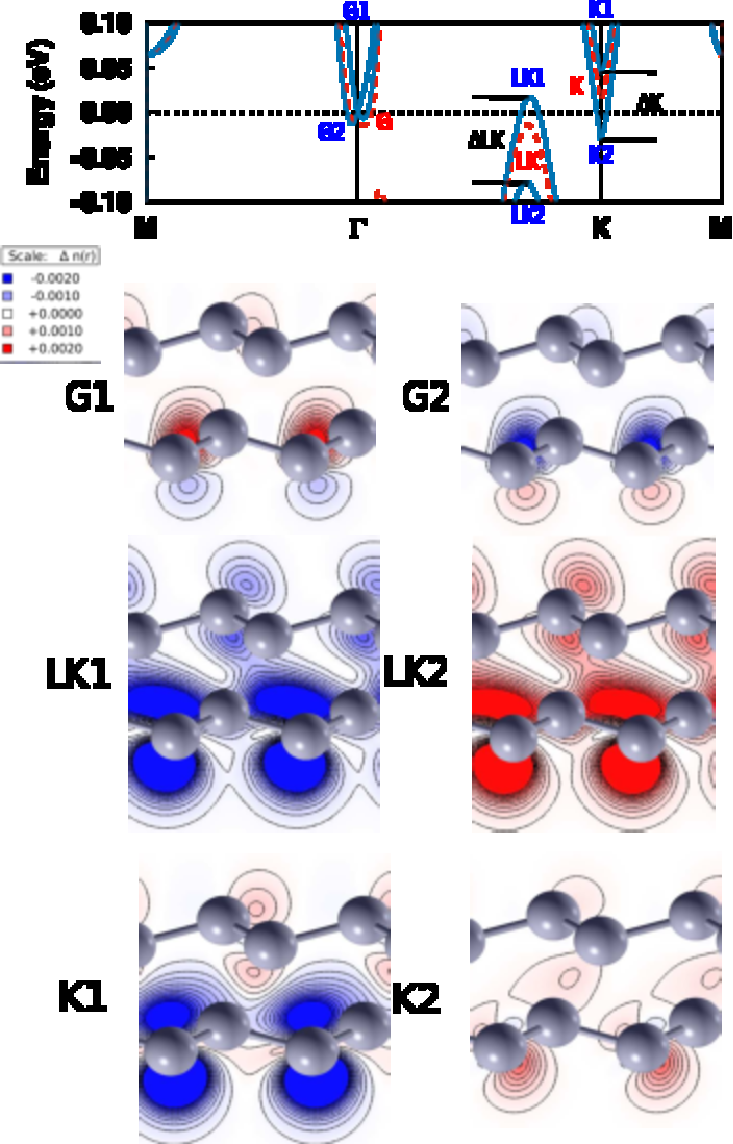}
	\caption{(color online) The enlarged band structure of the AA configuration with the angle of 5 degrees and a
        spacing of 3.4~\AA~between the two layers.
	 	The dotted red lines represent the bands without SOC and the blue lines represent bands with SOC.
	 	Bottom plots are the band decomposed charge densities at G1, G2, LK1, LK2, K1 and K2 k-points, respectively.
	 	From these charge density contour plots one can observe $\ket{\uparrow}$ and $\ket{\downarrow}$ states. It is to be noted that Fermi Energy is set to be at zero on y-axis.}
	\label{enlarged_aa_five}
\end{figure}

The figure \ref{enlarged_aa_five} is an enlarged version of the band structure of AA-\five~having an interlayer spacing
of 3.38~\AA.
From the non-SOC band structure represented in red dotted lines, a semimetallic nature is observed due to the interaction
between band at $\Gamma$ point and the band left to $K$ point.
Also, an indirect bandgap is observed between the band next to $K$ point and the band at $K$ point.
When the SOC interaction is taken into consideration, the former set of the bands overlap, while the latter indirect
bandgap is closed.
Let the band at $\Gamma$ point be labeled as ``G" before splitting and as ``G1" ``G2" for the top and bottom bands,
respectively, after splitting due to SOC.
Similarly, the bands corresponding to the $K$ point be labeled as ``K", ``K1" and ``K2" and the bands toward the left side
of the $K$ point are labeled as ``LK", ``LK1" and ``LK2", respectively.
The energy difference between the bands LK1 and LK2 is $\Delta$LK=95~meV and that between K1 and K2 is $\Delta$K=74~meV.
Clearly, the above effect is due to inclusion of SOC.

%...............................................................

%...............................................................

\newpage
We also plot the band decomposed charge density for those k-points as their characteristic is well documented and have
been used for the similar analyzes~\cite{MOS_SOC_SPLITING_1, MOS_SOC_SPLITING_2}.
By including the magnetization direction (of bands) and observations from band decomposed charge density, we note the
following: G1 has~$\ket{\uparrow}$ state, G2 has~$\ket{\downarrow}$ state, K1 has~$\ket{\downarrow }$ state, K2
has~$\ket{\uparrow}$ state, LK1 has~$\ket{\downarrow }$ state and LK2 has~$\ket{\uparrow}$ state.
Thus, in the direction from $LK$ to $\Gamma$ point the down channels overlap while the up channel has an indirect
bandgap of $\approx65$~meV.
It may also be noted that system only provides a conducting up~(down) channel and an indirect bandgap for the up~(down)
channel of the magnitude of 28.13~(49.29)~meV, in the direction of $K$ and $LK$ points.
We can thus conclude that there are distinct up and down channels.
Further, the splitting at $\Gamma$ point may seem to be Rashba-type splitting as the system has all the necessary
requirements such as lack of inversion symmetry, band splitting due to SOC and band splitting at $\Gamma$
point~\cite{Rashba_split}.
However, a charge density analysis show that it is not.
Interestingly, the characteristics of band structure is dominated by the bands at and near $K$ point.

\begin{figure}[H]
	\centering
	\includegraphics[scale=1.0]{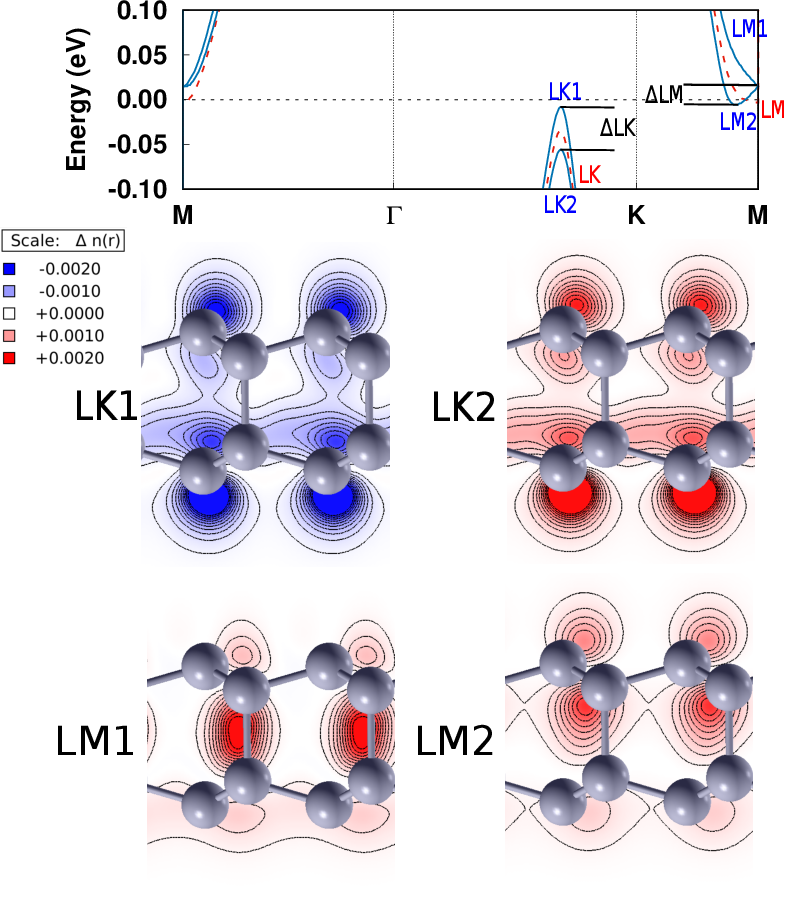}
	\caption{(color online) The enlarged band structure of the AB configuration with the angle of 70 degrees and an
         interlayer spacing of	$d=$4.6~\AA~between the two layers.
	 	The dotted red lines represent the bands without the SOC and the blue lines represent the bands with SOC.
	 	The bottom plots are the band decomposed charge densities at LK1, LK2, LM1 and LM2 k-points, respectively. It is to be noted that Fermi Energy is set to be at zero on y-axis.}
	\label{F7}
\end{figure}

The above-mentioned methodology can be used to interpret the band structure of the AB-\seventy~having
an interlayer spacing $d$=4.60~\AA.
From the band structure in figure \ref{F7}, an indirect bandgap between the band left to the $K$~point and band left to
$M$~point of the magnitude of 31~meV.
Due to the SOC, an indirect bandgap is reduced to null, while converting the semiconducting nature of the material into
semimetallic due to the splitting in both bands.
Using the same labeling convention as described earlier, the band left to $K$~point is labeled as ``LK" - before
splitting, ``LK1" and ``LK2" - after splitting with an energy difference between them being $\Delta$LK=48~meV.
Similarly, the band left to the $M$~point is labeled as ``LM", ``LM1" and ``LM2". having an energy difference between
LM1 and LM2 band is $\Delta$LM=22~meV.
From the band decomposed charge density (figure~\ref{F7}), along with the direction of magnetization of band we get,
LK1 as~$\ket{\downarrow}$ state, LK2 as ~$\ket{\uparrow}$ state and LM1 as~$\ket{\uparrow}$ state, LM2
as~$\ket{\uparrow}$ state.
This again leads to the AB-\seventy~structure having a distinct channel for up state with an indirect gap as 8.41~meV.
Hence, from the above observations for both structures, we can assert that the system above provides us with a distinct
up channel while the AA-\five~provides up as well as down channel.
Such a feature of band splitting has been observed in multilayer MoS$_2$ structure~\cite{MOS_SOC_SPLITING_2} and is
exploited for designing next generation of electronic devices leading to the development of
valleytronics~\cite{Valleytronics}, where the energy of an electron and simultaneously valley of the bands is used.

From charge density distribution of bands (Figure ~\ref{enlarged_aa_five} and~\ref{F7}) a spread between layers is
observed.
This spread of charge density can be directly related strength of the interlayer coupling.
As the charge density of these bilayer systems is well distributed in between both layers, the interlayer coupling can
be considered strong.
Additionally, to support that our system has a potential of being used in a valley device, the same procedure was used
at the $-K$ point (other valley).
For band structure and band decomposed charge density for the $-K$ valley, see supporting information~(SI:2, SI:3).
The calculated results are in an accord that our proposed system do have a merit of being used for valley device.

%...............................................................

%...............................................................

%...............................................................
\newpage
%...............................................................

\section{Conclusion}
We have studied the bilayer stanene with two stacking configurations AA and AB, using first principle density
functional theory.
By changing the interlayer spacing and stacking angles we found the minimum energy configuration of the bilayer.
For AB stacking configurations of angle 0$^0$ and 5$^0$, we observed two minima, while the other configurations
energies  were found to be of the same magnitude.
Hence, these systems do have a similar probability of being observed experimentally.
Nevertheless, we observe that a change in the stacking angle can affect the electronic properties of the system
significantly.
Further, by including spin-orbital coupling effects, we observe band splitting, which converts semimetallic nature of
the material into metallic.
A detailed analyzes of the band structure in conjunction with the band decomposed charge density as well as
magnetization directions of bands, we conclude that while the system AB-\seventy provides a distinct up channel
the AA-\five~system provides up as well as down channel, with indirect band gaps.
Our work indicates that the mechanically exfoliated bilayer stanene can be used for development of spin and valley physics
which can be used for next generation of electronic devices.

%...............................................................

\section{Acknowledgment}
The computational work described here is performed at the High Performance Computational Facility at IUAC, New Delhi,
India.
We would like to express our gratitude to them.
Also, we would like to thank the University Grant Commission of India for providing partial funding for the research work
through the UGC-BSR Research Startup Grant (Ref. No.F.30-309/2016(BSR)).

\bibliographystyle{unsrt}
\bibliography{REF}

\end{document}